\documentclass[10pt,journal,compsoc]{IEEEtran}

\ifCLASSOPTIONcompsoc
  \usepackage[nocompress]{cite}
\else
  \usepackage{cite}
\fi
\ifCLASSINFOpdf
\else
\fi

\pdfoutput=1\relax                   
\usepackage{graphicx}                
\DeclareGraphicsExtensions{.pdf,.png,.jpg,.jpeg} 

\graphicspath{{figures/}{pictures/}{images/}{./}} 

\usepackage{microtype}                 
\PassOptionsToPackage{warn}{textcomp}  
\usepackage{textcomp}                  
\usepackage{mathptmx}                  
\usepackage{times}                     
\usepackage{cite}                      
\usepackage{tabu}                      
\usepackage{booktabs}                  

\usepackage{xcolor}
\usepackage{xspace}
\usepackage{lipsum}
\usepackage{comment}
\usepackage{paralist}
\usepackage{hyperref}
\usepackage{enumerate}

\usepackage[inline]{enumitem}
\usepackage[noend]{algorithmic}
\usepackage{fontawesome}
\newcommand{\mzhihua}[1]{#1}

\newcommand{\xingbo}[1]{\textit{\textcolor{orange}{[Xingbo: #1]}}}

\newcommand{\ie}{i.e.}
\newcommand{\eg}{e.g.}
\newcommand{\etal}{\textit{et al.}}

\newcommand{\dq}[1]{``#1''}

\newcommand{\systemname}{{ShortcutLens}}
\newcommand{\name}{{\textit{ShortcutLens}}}
\newcommand{\vone}{Control Panel}
\newcommand{\vtwo}{Statistics View}
\newcommand{\vthree}{Template View}
\newcommand{\vfour}{Instance View}

\usepackage{algorithm}
\usepackage{algorithmic}
\usepackage{amsmath}
\usepackage{amsfonts}

\usepackage{lineno}

\begin{document}
%
\title{{\systemname}: A Visual Analytics Approach for Exploring Shortcuts in Natural Language Understanding Dataset}
%
%
%
%
\author{Zhihua Jin, Xingbo Wang, Furui Cheng, Chunhui Sun, Qun Liu, and Huamin Qu,~\IEEEmembership{Member,~IEEE}
\IEEEcompsocitemizethanks{\IEEEcompsocthanksitem Zhihua Jin, Xingbo Wang, Furui Cheng, and Huamin Qu are with  Hong Kong University of Science and Technology, Hong Kong, China. \protect\\
E-mail: \{zjinak, xwangeg, fchengaa, huamin\}@cse.ust.hk.
\IEEEcompsocthanksitem
 Chunhui Sun is with Peking University, Beijing, China. \protect\\
 E-mail: sch@pku.edu.cn.
\IEEEcompsocthanksitem
 Qun Liu is with Huawei Noah's Ark Lab, Hong Kong, China. \protect\\
 E-mail: qun.liu@huawei.com.
 }
\thanks{Manuscript received XX XX, 2022; revised XX XX, 2022.}}

%
%

\markboth{IEEE TRANSACTIONS ON VISUALIZATION AND COMPUTER GRAPHICS,~VOL.~XX, NO.~XX, XX~2022}%
{Shell \MakeLowercase{\textit{et al.}}: Bare Advanced Demo of IEEEtran.cls for IEEE Computer Society Journals}
%



\IEEEtitleabstractindextext{%
\begin{abstract}
Benchmark datasets play an important role in evaluating Natural Language Understanding (NLU) models. However, shortcuts---unwanted biases in the \mzhihua{benchmark datasets}---can damage the effectiveness of \mzhihua{benchmark datasets} in revealing models' real capabilities. Since shortcuts vary in coverage, productivity, and semantic meaning, it is challenging for NLU experts to systematically understand and avoid them when creating \mzhihua{benchmark datasets.} In this paper, we develop a visual analytics system, {\name}, to help NLU experts explore shortcuts in \mzhihua{NLU benchmark datasets.} The system allows users to conduct multi-level exploration of shortcuts. Specifically, Statistics View helps users grasp the statistics such as coverage and productivity of shortcuts in the \mzhihua{benchmark dataset.} Template View employs hierarchical and interpretable templates to summarize different types of shortcuts. Instance View allows users to check the corresponding instances covered by the shortcuts. We conduct case studies and expert interviews to evaluate the effectiveness and usability of the system. The results demonstrate that {\name} supports users in gaining a better understanding of \mzhihua{benchmark dataset} issues through shortcuts, inspiring them to create challenging and pertinent \mzhihua{benchmark datasets.}
\end{abstract}

}

\maketitle

\IEEEdisplaynontitleabstractindextext

%
\IEEEpeerreviewmaketitle

\ifCLASSOPTIONcompsoc
\IEEEraisesectionheading{\section{Introduction}\label{sec:introduction}}
\else
\section{Introduction}
\label{sec:introduction}
\fi

\begin{figure*}[htb]
\centering
  \includegraphics[width=\linewidth]{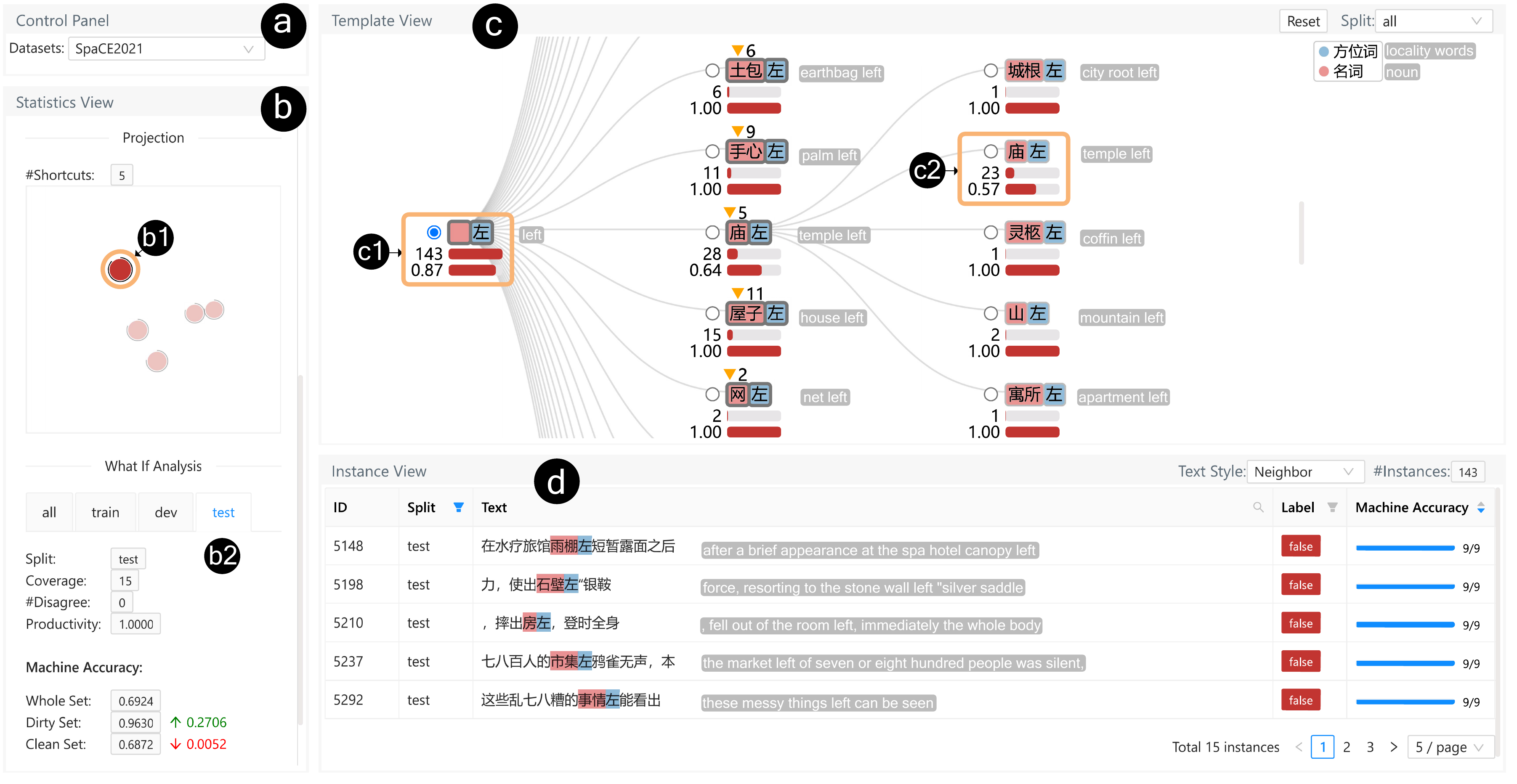}
  \caption{{\name} assists NLU experts in conducting the multi-level exploration of shortcuts in \mzhihua{NLU benchmark datasets.} (a) The {\vone} allows users to select the \mzhihua{benchmark dataset.} (b) The {\vtwo} helps users inspect the statistics about the \mzhihua{benchmark dataset} and shortcuts. It also allows users to conduct what-if analysis on shortcuts of interest. (c) The {\vthree} enables users to check the relationship of shortcuts and inspect the statistics about individual shortcuts. (d) The {\vfour} displays the instances covered by selected shortcuts from the {\vthree}.}
	\label{fig:teaser}
\end{figure*}

\IEEEPARstart{B}{enchmark} datasets play a fundamental role in Natural Language Understanding (NLU) by providing grounds for model evaluation and comparison~\cite{bowman2021will, ethayarajh2020utility}. Successful examples, like GLUE~\cite{wang2018glue} and SuperGLUE~\cite{wang2019superglue}, have been widely accepted in assessing models' NLU capability, which can benefit downstream applications, such as sentiment analysis~\cite{sun2019utilizing, xu2019bert} and fake news detection~\cite{kaliyar2021fakebert, jwa2019exbake}.

Building high-quality benchmark datasets is challenging. The effort is far beyond randomly collecting and labeling samples. Instead, researchers have to assess the quality of the benchmark dataset from different perspectives carefully and fix the data problems accordingly~\cite{paullada2021data}. 
\mzhihua{What is a high-quality benchmark dataset?
Except for the primarily used criteria, \eg, \textit{validity} (whether the dataset is correctly labeled) and \textit{diversity}, recent research argues that a high-quality \mzhihua{benchmark dataset} needs to be \textit{challenging} and \textit{pertinent}~\cite{kiela2021dynabench, bowman2021will}.} \textit{Challenging} means that the benchmark dataset should be able to reveal the gaps between different models~\cite{kiela2021dynabench}. For example, we do not want simple models to achieve human-level performance in the benchmark datasets easily.
\mzhihua{\textit{Pertinent}} indicates that the model's performance on the benchmark dataset should reflect its capability in the target task.
Intuitively, an NLU model should leverage the task-related words and language structures to make predictions instead of spurious biases.


\mzhihua{Benchmark datasets collected from the traditional data collection process~\cite{paullada2021data} cannot guarantee such criteria.
According to recent literature, shortcuts--unwanted biases in the datasets--undermine the quality of NLU datasets. Shortcuts are easily captured by the machine learning (ML) models in making predictions~\cite{geirhos2020shortcut}, allowing even simple models to achieve good performance in challenging prediction tasks. In NLU benchmark datasets, shortcuts can be words, phrases, and language structures that occur with different frequencies in different categories of sentence instances. Such shortcuts should be identified and removed to ensure the quality of the NLU benchmark dataset.}

\mzhihua{Most existing approaches target building \textit{challenging} datasets with fully-automated methods, such as adversarial filtering~\cite{sakaguchi2020winogrande, le2020adversarial} and counterfactual augmentation~\cite{niven2019probing,kiela2021dynabench}. 
However, these methods do not consider the \textit{pertinence} of the dataset simultaneously~\cite{branco2021shortcutted, bowman2021will}. 
Simple but representative instances may be removed from the original datasets. And the newly constructed datasets may still contain task-irrelevant patterns that can be obtained by the models in making predictions.
An alternative approach is to keep the experts in the dataset exploration and correction loop, where experts are informed of the potential shortcuts and decide whether to fix them according to their expertise. However, it is challenging for dataset creators to inspect and mitigate the shortcuts in the \mzhihua{benchmark dataset} by directly analyzing the instances one by one, which requires many human labor efforts. Interactive and efficient tools are desired, such as visualization tools, to explore shortcuts and gain insights into how to mitigate the shortcuts in the \mzhihua{benchmark dataset.}
}

In this paper, we collaborate with two NLU experts for four months and summarize a list of requirements for system design and develop the system called {\name}, which can help dataset creators conduct a multi-level exploration of the shortcuts in \mzhihua{NLU benchmark datasets,} as shown in Fig.~\ref{fig:teaser}. {\name} includes three views: {\vtwo}, {\vthree}, and {\vfour}. The {\vtwo} displays necessary statistics about the \mzhihua{benchmark dataset} and the potential shortcuts within. Users can also select a group of potential shortcuts and conduct the what-if analysis (Fig.~\ref{fig:teaser}(b)).
The {\vthree} employs hierarchical and interpretable templates to summarize the shortcuts. The glyph is proposed to visually represent the shortcuts (Fig.~\ref{fig:teaser}(c)).
The {\vfour} allows users to check the instances covered by selected shortcuts from the {\vthree} (Fig.~\ref{fig:teaser}(d)). 
Case studies and expert interviews are conducted to evaluate the effectiveness and usability of {\name}. 
The results demonstrate that {\name} can help dataset creators explore the shortcuts and gain insights into how to mitigate them in the \mzhihua{benchmark dataset.}


The contributions of this work can be summarized as follows:

\begin{itemize}
    \item A visual analytics system to help dataset creators systematically explore the shortcuts in \mzhihua{NLU benchmark datasets.}
    \item Case studies on understanding and mitigating shortcuts in \mzhihua{NLU benchmark datasets} and expert interviews to demonstrate the effectiveness and usability of {\name}.
\end{itemize}

\section{Related Work}
In this section, we introduce related work on criteria and guidelines of natural language understanding datasets construction, visualization for natural language processing, and visualization for improving dataset quality.


\subsection{Criteria and Guidelines of Natural Language Understanding Datasets Construction}
\label{sec:criteria_datasets}

In recent years, pre-trained language models like BERT~\cite{devlin2018bert} have achieved great success on NLU benchmarks, such as GLUE~\cite{wang2018glue} and SuperGLUE~\cite{wang2019superglue}. 
Although models achieve comparable and even superior performance to human performance on standard benchmarks, they can easily fail in challenging or real-world cases~\cite{kiela2021dynabench,wang2021adversarial,branco2021shortcutted}. 
It shows that these benchmarks are no longer adequate for evaluating human-like complicated and comprehensive language abilities.
One reason for such a phenomenon is that many benchmarks suffer from spurious biases that incentivize the inflating model performance.

According to prior research~\cite{poliak2018hypothesis,gururangan2018annotation,branco2021shortcutted}, many spurious biases in NLU benchmarks relate to shortcuts.
They can inflate the performance of models and mislead the researchers to improve it in the wrong direction. 
For example, Gururangan~\etal~\cite{gururangan2018annotation} found that the annotator is inspired by the examples in the annotation guideline to introduce artifacts into the \mzhihua{benchmark dataset} by simply substituting words. Therefore, words such as \dq{not} will be closely associated with specific labels, and such shortcuts can inflate the model performance. 
In addition, a line of research revealed that models tend to exploit simple functions or spurious statistical cues when performing NLU tasks~\cite{lai2021machine, mccoy2019right}.

Facing these issues, many works have proposed guidelines on dataset construction to measure and improve the benchmark quality. 
For example, clear and detailed documentation of \mzhihua{benchmark datasets} should be carefully considered~\cite{bender2018data,gebru2021datasheets}. They can help reveal the limitations of \mzhihua{benchmark datasets} and more precisely depict the properties of \mzhihua{benchmark datasets.} 
Kiela~\etal~\cite{kiela2021dynabench} encouraged dynamically collecting data and evaluating models to help mitigate the biases in the \mzhihua{benchmark datasets}.
Bowman~\etal~\cite{bowman2021will} proposed a more general guideline to help construct \mzhihua{benchmark datasets.} The benchmark should meet four criteria: (1) Good performance in the benchmark is an indicator of robust performance in the domain task. (2) The instances in the \mzhihua{benchmark datasets} should be accurately and unambiguously annotated. (3) It should be significantly difficult or large enough. (4) Social biases can be revealed through the benchmark and they should not be incentivized by the benchmark. 

However, these guidelines cannot be easily adopted in practice. Many criteria are coupled and impose great challenges for dataset creators to improve the dataset quality. For example, it is challenging to mitigate the bias in the \mzhihua{benchmark dataset} while preserving the capability of indicating the models' specific NLU ability.
To facilitate this process, we follow the guidelines to design our system to meet its desired properties. We mainly consider two important properties: The first one is to make \mzhihua{benchmark datasets} more challenging. The second one is to reflect the models' NLU capability instead of shortcuts, which means that it can imply robust performance.

\subsection{Visualization for Natural Language Processing}

There is a line of related research focusing on visualization for NLP models~\cite{choo2018visual}, including model understanding~\cite{strobelt2017lstmvis,ming2017understanding,derose2020attention,tenney2020language,DBLP:journals/tvcg/WangHJYWQ22}, debugging~\cite{laughlin2019visual,wu2020tempura}, and refinement~\cite{liu2018nlize,ming2019protosteer}. 
For NLP model understanding, Strobelt~\etal~\cite{strobelt2017lstmvis} proposed LSTMVis which can match hidden state changes with syntactic function changes in the text. Ming~\etal~\cite{ming2017understanding} developed RNNVis that employed the co-clustering technique to reveal the relationship between hidden states and words. 
Tenney~\etal~\cite{tenney2020language} further integrated different methods, such as projection and counterfactual generation, to assist model developers in investigating their models. 
For NLP model debugging, Laughlin~\etal~\cite{laughlin2019visual} utilized adversarial text generation to help researchers identify model deficits.  However, their work cannot be utilized to summarize the potential shortcuts in the \mzhihua{benchmark dataset.}
Wu~\etal~\cite{wu2020tempura} proposed hierarchical structural templates to summarize the query dataset and enable model developers to find the error patterns of their models. One issue of their templates is that they are limited to short query datasets, which cannot be applied to long-sentence summarization.
For NLP model refinement, NLIZE~\cite{liu2018nlize} allows users to interactively refine models via the attention matrix. 
Ming~\etal~\cite{ming2019protosteer} built ProtoSteer, which can help domain experts update a sequence model by interactively adding, deleting, and revising prototypes.

However, most of \mzhihua{these works} mainly focus on improving the model performance, neglecting to improve the dataset quality. Moreover, they do not provide sufficient help for dataset creators to find flaws in the benchmark dataset and gain insights into how to improve it. 



\subsection{Visualization for Improving Dataset Quality}

A group of work focuses on improving image or tabular dataset quality although it neglects the textual data~\cite{yuan2021survey}.
They can be categorized into two classes: data anomaly detection~\cite{kandel2012profiler, lakshminarayanan2017simple,lee2017training,bors2019capturing,chen2020oodanalyzer} and missing values detection~\cite{alemzadeh2020visual}. 
In terms of data anomaly detection, Chen~\etal~\cite{chen2020oodanalyzer} proposed OoDAnalyzer which can help improve OoD detection accuracy with a grid-based visualization. 
For missing value detection, Alemzadeh~\etal~\cite{alemzadeh2020visual} developed VIVID 
to diagnose the root causes of missing values through multiple coordinated visualizations. 
\mzhihua{These works} cannot be easily adapted to textual data because textual data is unstructured and not easy to explore, while tabular data is structured data and image data is easy to understand through a glimpse. 
A recent study called DQI~\cite{mishra2020dqi} can help reveal shallow statistics of textual datasets. 
However, it is not effective for investigating \mzhihua{benchmark datasets} that are designed for evaluating the complicated reasoning ability of models, as models can exploit more complex shortcuts. 
Moreover, it does not integrate different types of statistics into compact visualizations, making the analysis inefficient.

Therefore, we aim to develop a tool to explore the shortcuts in \mzhihua{NLU benchmark datasets} to fill in the research gap. Current work fails to reveal the shortcut in a unified, hierarchical, and understandable way. Instead, our work adopts hierarchical and understandable templates to help dataset creators explore and identify potential shortcuts.

\section{Background}
\label{sec:background}
A wide range of tasks for evaluating the NLU capability is usually formulated as the single sentence classification task or multiple sentence classification task~\cite{wang2018glue, wang2019superglue}. In this paper, we mainly focus on the single sentence classification tasks. We will illustrate two examples, including spatial reasoning and grammatical acceptability classification.


\textbf{Spatial reasoning.} Spatial reasoning is a cognitive process to cope with environments, which requires constructing representations of spatial relationships and transformations between objects~\cite{clements1992geometry,mirzaee2021spartqa}. 
To test the model's spatial reasoning capability, a team of researchers recently created a new \mzhihua{benchmark dataset} called SpaCE2021 to evaluate the textual spatial reasoning capability of the models.  
An assumption is that if the model has spatial reasoning capabilities, then it must be not only able to recognize the correct spatial information but also able to recognize abnormal and incorrect spatial information. 
For example, for the sentence \dq{sign a name on all sides}, people can realize that it is weird because \dq{a name} is usually not signed on \dq{all sides}.  For the sentence \dq{walking under a train}, people can clearly know that in most cases, no one walks under a train. \mzhihua{Instead, the sentences \dq{sign a name on a book} and \dq{walking on the train} have correct spatial information.} 

\textbf{Grammatical acceptability judgment.} 
Warstadt~\etal~\cite{warstadt2019neural} built a \mzhihua{benchmark dataset} called CoLA to test models' ability on grammatical acceptability judgment. 
For example, the sentence \dq{This building is than that one.} is considered a grammatically unacceptable sentence, while a slightly modified sentence \dq{This building is taller and wider than that one.} will be regarded as a grammatically acceptable sentence. 

\section{Abstractions}
\label{sec:abstractions}

In this section, we introduce an abstraction of the domain problem, analysis workflows, and design requirements in NLU benchmark dataset construction. 

\mzhihua{We collaborated with a research team from a large company for four months in developing SpaCE2021, a Chinese spatial reasoning benchmark dataset. We closely worked with two team members (\textbf{E1}, \textbf{E2}), who are also our co-authors. \textbf{E1} is a senior Ph.D. student with more than two years of experience in NLU dataset construction. \textbf{E2} is the team leader with more than twenty years of experience in developing NLP models and benchmark datasets in commonsense reasoning and machine translation. Through our observation of their dataset construction and validation process and multiple rounds of interviews, we characterize the problems, their workflow, and requirements for designing an interactive tool as follows.}

\subsection{Domain Problem}
\label{sec:domain_problem}
\mzhihua{The benchmark dataset development consists of three stages: \textit{data collection}, \textit{augmentation or filtering}, and \textit{validation}. Taking SpaCE2021 as an example, experts (including both NLP experts and linguists) first collected sentences with spatial relationship descriptions from novels, essays, and fiction. 
Then they randomly replaced some locality words like ``inside'' with other locality words like ``outside'' in the sentence. They further distributed the modified sentences to a group of annotators to judge whether the spatial information in the modified sentences is correct or not.
In the validation stage, they mainly focused on checking the annotator agreement on the labels and fixing data contamination issues.
However, when applying this dataset for benchmarking, the experts found that simple models achieve human-level performance, which was unreasonable considering the difficulty of the prediction problems. Thus, they began to notice the shortcut problems in the benchmark dataset and expected to explore and summarize the shortcuts to gain actionable insights to improve the benchmark dataset quality.} 

\subsection{Workflow}
\mzhihua{
The experts used to find shortcuts based on their intuitions. They first inspect some challenging but correctly solved instances. From these instances, they summarize a set of words (usually nouns) that they think to be shortcuts. Then the experts test the frequency of these words in different categories of samples (\ie, true and false samples for a binary classification problem). When a word's distribution is biased in different categories, it is likely to be a shortcut. The experts will record these words and adjust the datasets (\eg, removing the related instances) to test whether the dataset's quality is improved.} \mzhihua{However, the existing workflow only considers narrow-scoped shortcuts and heavily relies on experts' intuitions. To address these limitations, we need an interactive visual analytics tool to extract and organize those shortcuts in an interpretable, hierarchical, and unified way.}

\subsection{Design Requirements}

We summarize the requirements of designing an interactive visual analytics tool for exploring shortcuts in the NLU benchmark dataset.

\label{subsec.design_requirements}
\textbf{R1: Provide an overview of shortcuts.} 
\mzhihua{An overview helps experts make sense of the potential shortcuts. 
The experts commented that certain statistical information is critical for them to understand the importance of each shortcut, including the number of instances covered by the potential shortcuts (i.e., coverage), the percentage of the correct predicted instances in covered instances (i.e., productivity)~\cite{niven2019probing, branco2021shortcutted, wu2020tempura, mishra2020dqi}. 
Besides, the experts suggested that the shortcuts should be organized into a hierarchy.}
After checking the overview of potential shortcuts, they may select a group of potential shortcuts, and they further inspect them to understand the details behind them.

\textbf{R2: Associate shortcuts with corresponding instances.}
\mzhihua{Experts are willing to inspect individual shortcuts and quickly find their corresponding instances. They commented that it is essential for them to ``\textit{find concrete examples}'' and ``\textit{understand the shortcuts based on the context}''. 
Besides, they expect the system to highlight the words related to the shortcuts in the instances.}

\textbf{R3: Estimate the effects of removing shortcuts.} 
\mzhihua{After exploring and identifying shortcuts in datasets, the next step is to fix them, \eg, removing the corresponding instances~\cite{laughlin2019visual, kiela2021dynabench,wang2021adversarial,bowman2021will}.
However, removing these instances will sometimes bring other shortcuts, which cannot necessarily improve the dataset quality. 
Besides, conducting a formal test (including training large language models) can be time-consuming.
Instead, experts expect first to estimate the influence of mitigating them. 
For example, suppose users want to remove specific instances covered by specific shortcuts. In that case, it is helpful to know the changes in the machine accuracy between the original set and the set after cleaning instances.}
Such functionality can help dataset creators to decide the priority of fixing those shortcuts.
\section{\systemname}

Based on the derived design requirements, we design {\name} to support dataset creators in investigating potential shortcuts in \mzhihua{NLU benchmark datasets.} 
\mzhihua{In this section, we introduce an overview of the system first. Then we introduce the scope and relationship of shortcuts and what-if analysis. Design choices of individual views are introduced next. Finally, we introduce the shortcut mining and aggregation algorithm.}

\subsection{System Overview}

The system consists of three major modules: storage, shortcut mining module, and interactive visualization module. The storage module mainly manages the \mzhihua{NLU benchmark datasets.} The shortcut mining module is used to extract the potential shortcuts and parse the text for each instance. The interactive visualization module allows users to interact with the system to explore the potential shortcuts and conduct the what-if analysis. The storage and shortcut mining modules are built upon Python and then are integrated into a backend server built upon the Flask. Regarding the Chinese \mzhihua{NLU benchmark dataset,} we use the HanLP\footnote[1]{https://www.hanlp.com/} to extract the Part-of-Speech (POS)\footnote[2]{https://hanlp.hankcs.com/docs/annotations/pos/pku.html} and embedding of each word in the instances. For the English \mzhihua{NLU benchmark dataset,} we use spaCy\footnote[3]{https://spacy.io/} to extract the POS\footnote[4]{https://universaldependencies.org/u/pos/} and word embedding. The interactive visualization module is implemented as the frontend supported with browser using React, Typescript, and D3. 

\subsection{Scope and Relationship of Shortcuts}
To help extract the potential shortcuts in the dataset, it is desirable to have a suite of algorithms to mine the potential shortcuts from the \mzhihua{benchmark dataset.} 
We first assume that the potential shortcuts are matching-based shortcuts, that is, if the text contains the specific pattern, it will be considered covered by these potential shortcuts. 
Inspired by the defined shortcuts in DQI~\cite{mishra2020dqi} and the structural templates which include POS and \mzhihua{named entities} from Tempura~\cite{wu2020tempura}, we define the pattern as a set of spatially-related words. 
Specifically, we consider one word with its POS, or two words with their POS and their relative position (\ie, how many words are in between) to define the pattern or template. For example, we consider one pattern as follows: (1) The first word \dq{The} is a determiner. (2) The second word \dq{will} is a verb. (3) The first word is in front of the second word, and there is a word between them. Such a pattern can match sentences such as \dq{The men will all leave.}
The components of the potential shortcuts can be abbreviated, that is, some of their components can appear optionally in the potential shortcuts. For example, in the template above, the scope of the second word can be extended by considering this word as a verb, and not just the specific word \dq{will}. The newly defined template is the parent of the original template.
The mining algorithm selects shortcuts with high \textit{productivity} and \textit{coverage}~\cite{niven2019probing,branco2021shortcutted}. 
Coverage refers to the number of instances covered by the shortcut. The prediction label of the potential shortcut is defined as the dominant label in the label distribution. Productivity is calculated as the percentage of covered instances with the same label as the prediction label in covered instances. \mzhihua{Details of the shortcut mining algorithm are illustrated in Sec.~\ref{sec:shortcut_mining}.}

After running the shortcut mining algorithm, we find that there are a large number of potential shortcuts in the \mzhihua{benchmark dataset.} 
One of the reasons is that the number of different words is quite large. It imposes difficulty for users to explore such large-scale potential shortcuts. 
Therefore, we aggregate similar potential shortcuts based on the semantic meaning of the words to reduce the difficulty for users to explore.
\mzhihua{Details of the shortcut aggregation algorithm are illustrated in Sec.~\ref{sec:shortcut_aggregation}.}

\subsection{What-If Analysis}

In order to help dataset creators decide which potential shortcuts should be fixed and what actions should be taken to fix the potential shortcuts, a what-if analysis on shortcuts of interest is necessary for dataset creators to make decisions. 
Possible actions include constructing new instances, modifying the existing instances, or removing the existing instances to battle against a group of potential shortcuts. 
However, after taking action, it is hard for dataset creators to know whether it really resolves the issues. Therefore, a necessary what-if analysis should be considered. We will take an initial attempt to consider the influence of the removal of a group of instances.

After users have focused on a group of shortcuts of interest, the instances covered by these shortcuts will be found. We will consider the instances covered by them as the dirty set and the instances not covered by them as the clean set. 
When calculating the productivity for a group of potential shortcuts, it is possible that one instance is covered by multiple potential shortcuts with different prediction labels. Those instances will be considered as \dq{disagreed} instances. The number of \dq{disagreed} instances will be calculated. Then, the productivity of a group of shortcuts is defined as the percentage of covered \dq{agreed} instances with the same prediction label as the corresponding potential shortcuts. Such metrics indicate the impact of those potential shortcuts.

\mzhihua{If the machine prediction results are provided by the users, we will calculate the machine accuracy for the dirty set and clean set, respectively.} It is interesting to compare the machine accuracy between the whole set and the clean set to see whether the removal of potential shortcuts leads to the degrading of the performance metrics in the clean set. Such indicators can help dataset creators understand the potential benefits gained from the removal of the instances covered by those potential shortcuts.

\subsection{Visualization}
The {\name} visualization module consists of {\vone}, {\vtwo}, {\vthree}, and {\vfour}. The {\vone} allows users to select the \mzhihua{benchmark dataset} to explore the potential shortcuts (Fig.~\ref{fig:teaser}(a)). The {\vtwo} assists dataset creators in inspecting the statistics about the \mzhihua{benchmark dataset,} the potential shortcuts, and baseline models provided by the users (Fig.~\ref{fig:teaser}(b)). It also allows users to select a group of potential shortcuts to conduct what-if analysis and further inspect in the {\vthree}. The {\vthree} supports dataset creators in inspecting the individual potential shortcuts and the relationship between the potential shortcuts (Fig.~\ref{fig:teaser}(c)). After users select one potential shortcut in the {\vthree}, the {\vfour} displays the corresponding instances for dataset creators to further inspect (Fig.~\ref{fig:teaser}(d)). Users can choose to check the neighborhood of the highlighted matched text, which can quickly determine the location and context of the potential shortcut.

\begin{figure}[htb]
\centering 
\includegraphics[width=0.9\linewidth]{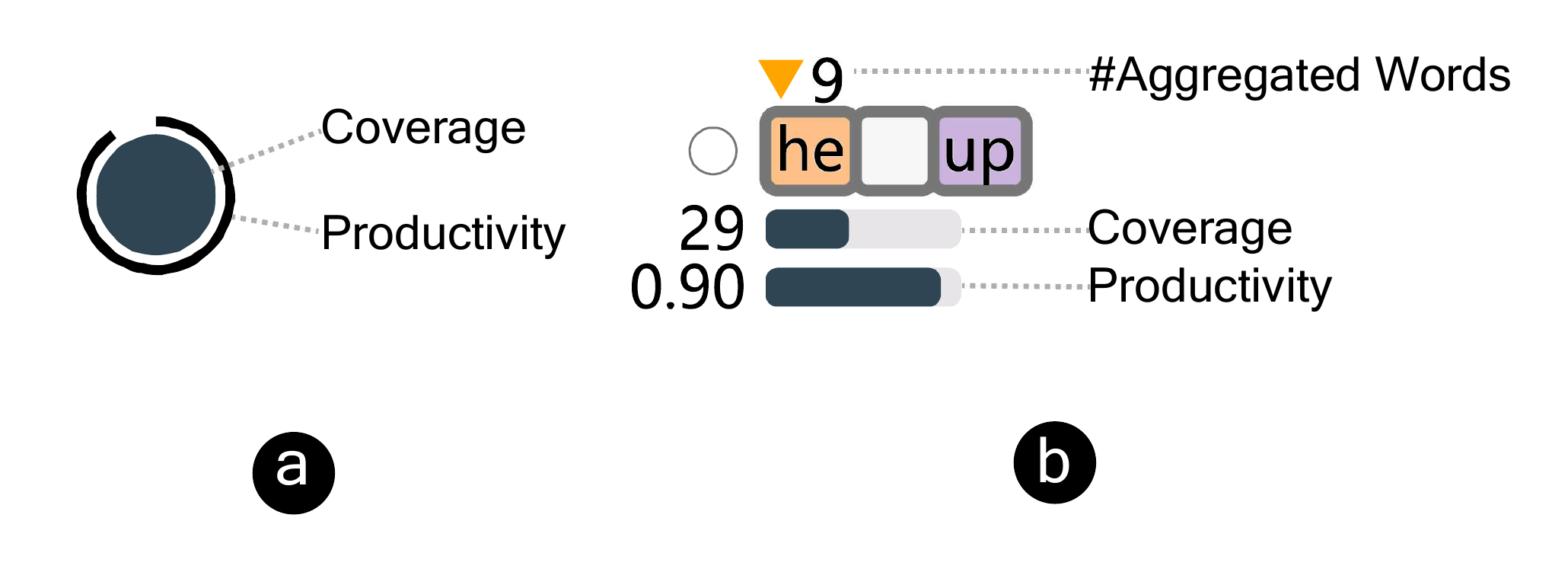}
\caption{The glyphs are used to represent the shortcuts. (a) The {\vtwo} uses a circle-based glyph to encode the productivity, coverage, and prediction label of the shortcut. (b) The {\vthree} uses a block-based glyph to encode the template of the shortcut. Coverage, productivity, and the number of aggregated words of the shortcuts will be displayed accordingly. } 
\label{Fig.glyph_illustration}
\end{figure}
\subsubsection{Statistics View}
The {\vtwo} helps dataset creators gain an overview of the \mzhihua{benchmark dataset,} potential shortcuts, and baseline models, as shown in Fig.~\ref{fig:teaser}(b) (\textbf{R1}). It consists of three panels, Instances, Shortcuts, and Machine Accuracy. The Instances panel displays the statistical summary of the \mzhihua{benchmark dataset.} It shows the number of instances and the label distribution in the whole \mzhihua{benchmark dataset} and different sets of the \mzhihua{benchmark datasets.} The Machine Accuracy panel shows the accuracy of individual models on the specific set. The Shortcuts panel allows users to explore the shortcuts in the Projection plane. 
Users can filter the shortcuts by productivity and coverage in the Filters plane.

\textbf{Visual design.} We use a circle-based glyph to represent the potential shortcuts and encode the relevant metrics about the potential shortcuts, as shown in Fig.~\ref{Fig.glyph_illustration}(a).
The radius of the circle encodes the coverage of the shortcut. The outer ring of the circle encodes the productivity of the shortcut. The color of the circle encodes the prediction label of the shortcut. For the binary classification task, we use red color to encode the label \textit{\dq{false}} and dark blue color to encode the label \textit{\dq{true}}.  The experts said that shortcuts with similar properties should be displayed in a near region, and it will be useful for users to inspect them. 
Thus, we use the UMAP~\cite{mcinnes2018umap} projection algorithm with collision avoidance to calculate the layout of the glyph of potential shortcuts. 
The distance between different potential shortcuts $a$ and $b$ is calculated as follows:

\begin{equation}
\begin{split}
Dist(a,b)&=|Prod_a-Prod_b|^2+|Norm(Cover_a)-Norm(Cover_b)|^2
\\& +\mathbb{I}\{Pred_{a}\ne Pred_{b}\}
\end{split}
\end{equation}

where $Prod$ is the productivity of the potential shortcut, $Cover$ is the coverage of the potential shortcut, $Norm(*)$ is the normalization function, and $Pred$ is the prediction label of the potential shortcut. $\mathbb{I\{*\}}$ is an indicator function. If the expression is true, the value is one. Otherwise, it is zero. 


\mzhihua{We considered one design alternative of circle-based glyphs. We can use two bars to encode the productivity and coverage of the shortcut and the color to encode the prediction label of the shortcut. However, the experts commented that it is more intuitive to use arc angles to present the percentage (i.e., productivity). Therefore, we decided to use a circle-based glyph for showing productivity and coverage, where coverage is encoded by circle radius, the angle of the outer arc represents productivity, and the circle color indicates the prediction label.}



\textbf{Interaction.} After lasso-selecting a group of potential shortcuts, the {\vthree} will display the corresponding potential shortcuts. Moreover, the what-if statistics will be displayed in the following spaces. It will demonstrate the coverage and productivity of the selected potential shortcuts. If the machine prediction results are provided by users, it will also display the difference in machine accuracy in the clean set or the dirty set. It can help dataset creators to estimate the influence of removing the instances covered by those shortcuts (\textbf{R3}).





\subsubsection{Template View}
To allow dataset creators to further inspect the selected potential shortcuts and the relationship between the potential shortcuts (\textbf{R1}), the {\vthree} is designed to display more fine-grained information of potential shortcuts ordered in a hierarchical structure (Fig.~\ref{fig:teaser}(c)).

\textbf{Visual design.} The experts commented that a vivid representation of the shortcuts is desirable for an intuitive understanding of the semantic meaning of the shortcuts.
Therefore, we use a block-based glyph to encode the matching content of the potential shortcuts, as shown in Fig.~\ref{Fig.glyph_illustration}(b). Each block represents a word. If the background color of the block is white, it means it will match nothing but stands as the placeholder for the relative position. The other background color of the block encodes the POS. The words in the block are the specified matched words. For example, as shown in Fig.~\ref{Fig.glyph_illustration}(b), it will match sentences with the pattern where the first word is the adposition \dq{up}, the second word in front of the first word is a pronoun which is similar to \dq{he}, and there is a word between those two words. The coverage and productivity of that potential shortcut are individually encoded by the bar width. If the shortcut is generated via the aggregation of similar words, it will have a triangle upon the representative of the word sets. Also, the number of aggregated words is displayed near the triangle. Such a glyph can help users understand the template behind the potential shortcuts. 

\textbf{Interaction.} If users change the split, the coverage and productivity of the shortcut will only be calculated on the selected split. 
Users can click the shortcuts with thicker borders of the blocks to expand the children of that potential shortcuts to check more fine-grained information about that potential shortcut. Moreover, the relationship between the shortcuts is also displayed. Users can also click the radio button in one of the potential shortcuts to further inspect the instances covered by it in the {\vfour} (\textbf{R2}). 




\subsubsection{Instance View}
To help dataset creators understand the template presented in the {\vthree}, as shown in Fig.~\ref{fig:teaser}(d) (\textbf{R2}), users can explore the covered instances in the {\vfour}. The experts commented that it is better to check the neighborhood of the matched text. 
\mzhihua{Thus, users can select the text style \dq{Neighbor} to show only the neighborhood of the matched text if the text is too long to inspect. The three words to the left and three words to the right of the highlighted text are considered the neighborhood of the matched text.} 
If the text is short, users can switch the text style to \dq{Full} to check all the text in the instances. The matched text will be highlighted. The background color of matched text depends on the POS of individual words, which has the same encoding in the {\vthree}. The table also displays the split, label, and machine accuracy for users to inspect. Users can search text, filter instances based on split or label, and sort the machine accuracy in the table to find the instances of interest.

\subsection{Shortcut Mining and Aggregation}
\begin{figure}[htb]
\centering 
\includegraphics[width=\linewidth]{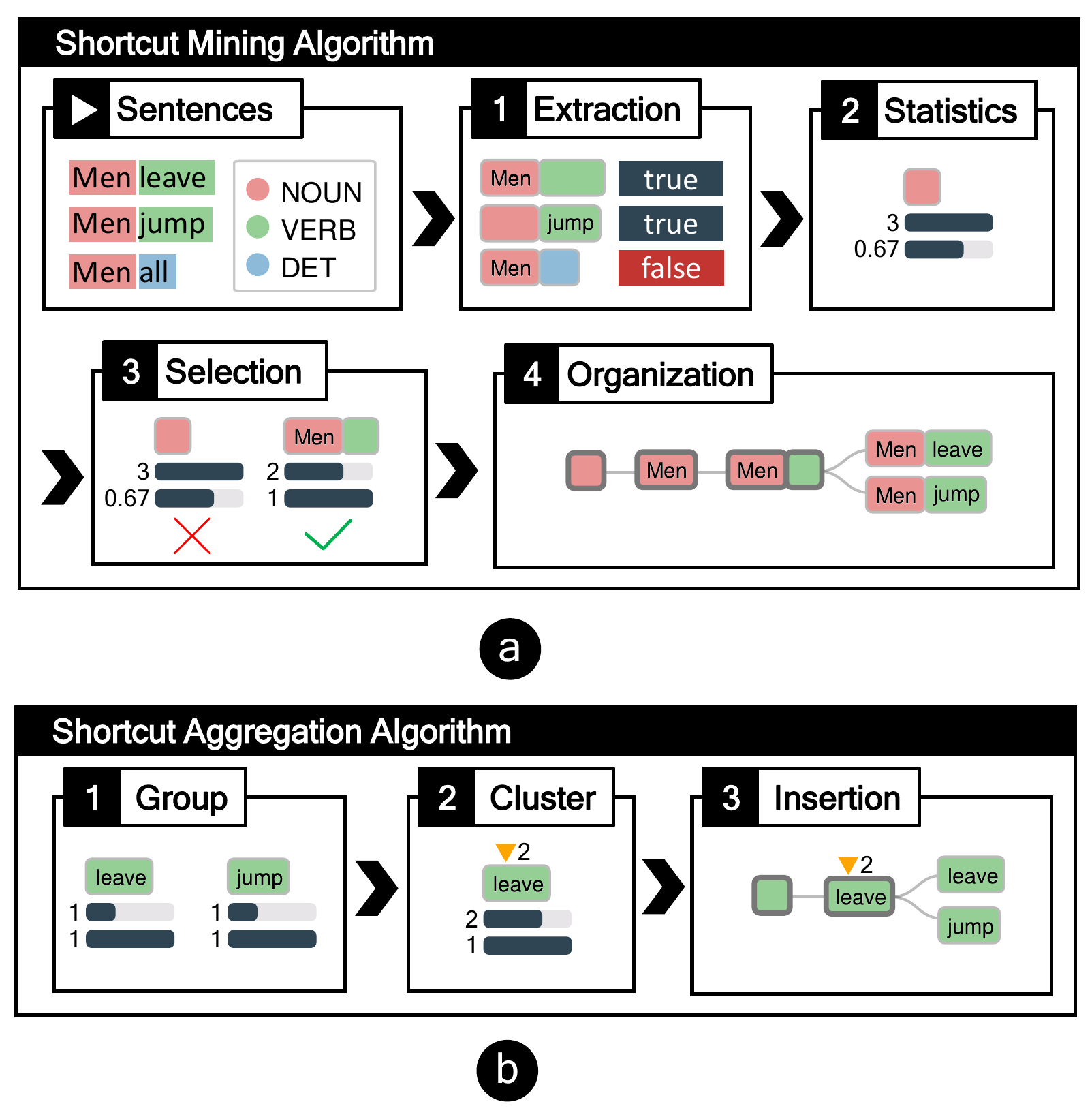}
\caption{\mzhihua{Shortcut mining algorithm (a) is to extract, select, and organize shortcuts. Shortcut aggregation algorithm (b) is to cluster shortcuts and insert representative shortcuts of clusters into the hierarchy.} } 
\label{Fig.methods}
\end{figure}

\mzhihua{Shortcut mining and aggregation algorithm will be introduced in the following paragraphs. The overview of workflow of those algorithms is depicted in Fig.~\ref{Fig.methods}.}

\subsubsection{Shortcut Mining}
\label{sec:shortcut_mining}
The basic idea of the mining algorithm (Fig.~\ref{Fig.methods}(a)) is extracting potential shortcuts via an exhaustive method. Then it will filter potential shortcuts based on the productivity and coverage. Finally, it will organize the shortcuts into a hierarchical structure. The details of each step in the algorithm are illustrated as follows: 

\textbf{Step 1: Extract Shortcuts.} Once we have the POS and words for a specific text, we will enumerate each word in the text to determine the POS of the first word and then its specific word. We will enumerate the possible relative position to get the position of the next word and its specific POS and word. 

\textbf{Step 2: Calculate Shortcuts Statistics.} The label distribution for each potential shortcut will be updated accordingly. If the specific instance has been considered in the potential shortcut, we will only consider that the potential shortcut covers that instance once. Then, we will calculate the coverage and productivity for each potential shortcut. 

\textbf{Step 3: Select Shortcuts.} We will filter the potential shortcuts by a preset minimum coverage and productivity in the whole set. Also, we will consider the minimum coverage and productivity in each set, like the training set, development set, and test set. 

\textbf{Step 4: Organize Hierarchy of Shortcuts.} We will organize the filtered potential shortcuts into a hierarchical structure based on the definition of the parent relationship. To reveal the relationship between the filtered potential shortcuts, we will track the paths from the root of the hierarchy of the potential shortcuts to filtered potential shortcuts and the children of filtered potential shortcuts.

\subsubsection{Shortcut Aggregation}
\label{sec:shortcut_aggregation}
The basic idea of the shortcut aggregation algorithm (Fig.~\ref{Fig.methods}(b)) is using hierarchical clustering to aggregate potential shortcuts with similar words based on word embedding into a new shortcut and insert this shortcut into the original hierarchy of shortcuts. The details are illustrated as follows: 

\textbf{Step 1: Group Mergeable Shortcuts and Calculate Distance Matrix.} We first check whether the two potential shortcuts can be merged, and then we calculate the distances between every two potential shortcuts. If two potential shortcuts share the same parent, the same prediction label, and the final component of each potential shortcut is the word, we will calculate the distances for the two final words. The distance between two final words $ftok$ is defined as $1-Sim(ftok_{a}, ftok_{b})$. The similarity between two words $Sim$ is defined as the cosine similarity of word embeddings for those two words. Then the distances between two potential shortcuts will be set to the distances between two final words. If two potential shortcuts do not satisfy the conditions for merging, the distances between these two potential shortcuts will be set to infinity. 

\textbf{Step 2: Cluster Shortcuts.} We adopt hierarchical clustering with complete linkage to cluster similar potential shortcuts based on the distances between two potential shortcuts~\cite{clusterAnalysis}. \mzhihua{We have empirically set the clustering methods as the complete linkage as we found that such a method can ensure that the distance between each shortcut will be under a certain threshold, which is empirically set as 0.75.} 

\textbf{Step 3: Insert Representative Shortcuts into the Hierarchy.} If a cluster of potential shortcuts has more than one potential shortcut, it will be converted into a new potential shortcut, and this new potential shortcut will be inserted into the hierarchy of the potential shortcuts. 
The final word of this new potential shortcut is defined as the final word of potential shortcuts with the smallest average distances to all other final words in the same cluster. This new potential shortcut covers the instances covered by all potential shortcuts in that cluster. 

\section{Evaluation}
In this section, we report the case studies with two experts and the results of expert interviews to demonstrate the effectiveness and usability of our system. \mzhihua{Actionable insights in case studies are verified in Appendix A.}

\begin{figure}[htb]
\centering 
\includegraphics[width=\linewidth]{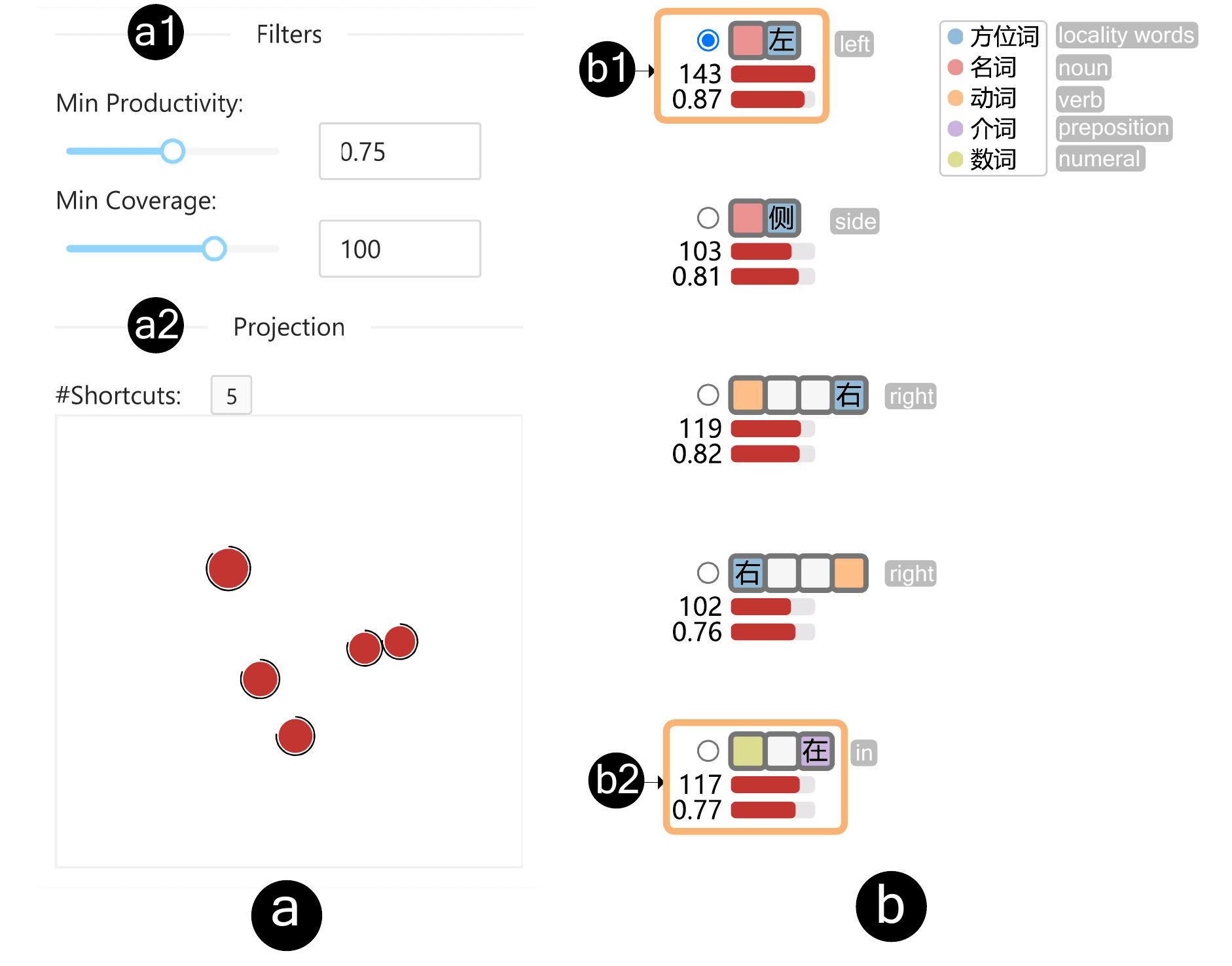}
\caption{E1 set the filters in the Statistics View (a) and five shortcuts satisfy the conditions. E1 found that four shortcuts consist of locality words while one shortcut consists of other kinds of words in the {\vthree} (b).} 
\label{Fig.case_study_one_fig_one}
\end{figure}
\subsection{Case-I: SpaCE2021 Dataset}
\label{sec:case_one}

\begin{figure*}[htb]
\centering 
\includegraphics[width=\linewidth]{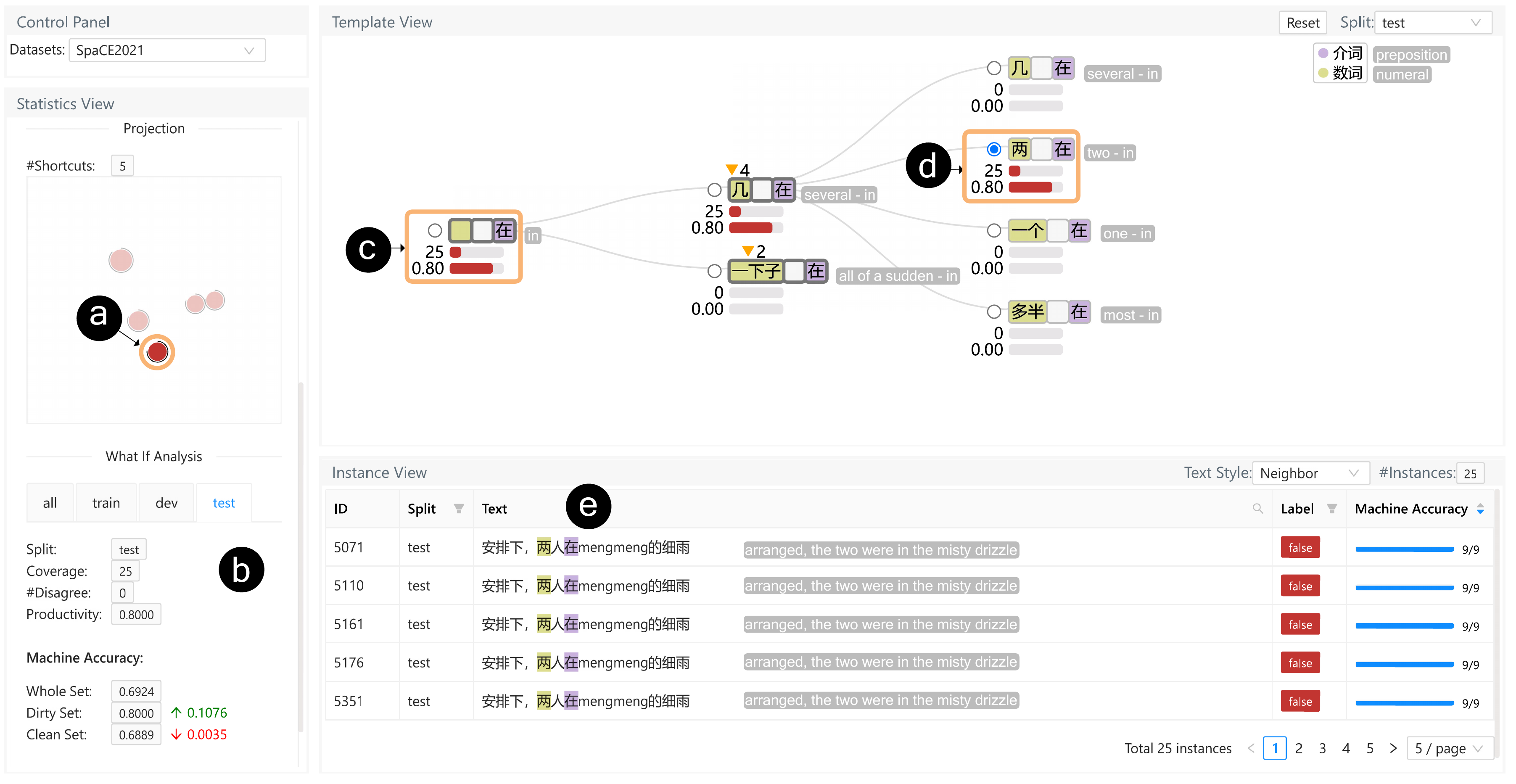}
\caption{E1 chose a shortcut in the Projection plane (a). The productivity of this shortcut is similar to machine accuracy in the dirty set in the What-If Analysis plane of the {\vtwo} (b). E1 found that only one child of that shortcut (c) covers the test instances (d). Most of the instances are with similar neighborhoods around the highlighted text (e).} 
\label{Fig.case_study_one_fig_three}
\end{figure*}
We conducted this case study with the dataset creator (E1) on spatial reasoning in Chinese. \mzhihua{The background of E1 is depicted in Sec.~\ref{sec:abstractions}.} The team of E1 has made an initial attempt at constructing the \mzhihua{benchmark dataset} for SpaCE2021. 
\mzhihua{The dataset description and collection process are depicted in Sec.~\ref{sec:background} and Sec.~\ref{sec:domain_problem} respectively.} 
After training nine models on this \mzhihua{benchmark dataset,} he found that the machine performance is higher than expected.
He decided to further inspect whether the \mzhihua{benchmark dataset} is good enough to evaluate the models' capability in spatial reasoning. 
He was wondering whether the \mzhihua{benchmark dataset} suffers from the shortcuts.

\mzhihua{E1 selected the SpaCE2021 benchmark dataset in the {\vone} (Fig.\ref{fig:teaser}(a)). In the {\vtwo}, E1 found that the percentage of instances with the label \textit{\dq{false}} in the test set is 0.5416 while the average machine accuracy on the test set is 0.6924, which is significantly better than the majority baseline on the test set. Given the situation that the model performance is quite good, E1 decided to further explore the potential shortcuts in the benchmark dataset. E1 set the minimum productivity as 0.75, which surpasses the average model accuracy, and the minimum coverage as 100 to filter the potential shortcuts (Fig.~\ref{Fig.case_study_one_fig_one}(a1)). There are five potential shortcuts satisfying this condition as shown in Fig.~\ref{Fig.case_study_one_fig_one}(a2). E1 further checked the What-If Analysis plane. The productivity of those shortcuts is 0.8136. In terms of machine accuracy, the performance on the dirty set is 0.0807 higher than the performance on the whole test set. E1 thought that those potential shortcuts can have a high impact on the model performance.}
E1 found that four potential shortcuts consist of locality words but one of them consists of other kinds of words (Fig.~\ref{Fig.case_study_one_fig_one}(b)). 
Since spatial reasoning requires a deep understanding of locality words in the sentence, it is more interesting to see that there exist potential shortcuts with no locality words, as shown in Fig.~\ref{Fig.case_study_one_fig_one}(b2). To understand this phenomenon, E1 decided to further explore one of the potential shortcuts with locality words (Fig.~\ref{Fig.case_study_one_fig_one}(b1)) and the shortcut without locality words (Fig.~\ref{Fig.case_study_one_fig_one}(b2)) (\textbf{R1}, \textbf{R3}). 

\subsubsection{Analyzing Shortcuts with Locality Words}
E1 selected the shortcut with the locality word \dq{left} with noun word ahead in the Projection plane (Fig.~\ref{fig:teaser}(b1), Fig.~\ref{fig:teaser}(c1)) and checked the What-If Analysis plane (Fig.~\ref{fig:teaser}(b2)). E1 found that this potential shortcut covers 15 instances in the test set \mzhihua{with productivity of 1.} In the dirty set, which includes instances covered by this shortcut, the machine accuracy \mzhihua{achieves 0.963.} It indicates that the models may find that this kind of instance is easier to answer and tend to exploit this kind of shortcut.  E1 further checked the Template View and expanded the children of that potential shortcut (Fig.~\ref{fig:teaser}(c1)). E1 found that there exist some children shortcuts with low productivity, like \dq{temple left}, which indicates that there exist some cases battling against this potential shortcut (Fig.~\ref{fig:teaser}(c2)). However, this potential shortcut covers 143 instances in total and has high productivity of 0.87, which means that other children of this shortcut tend to have high productivity. E1 further checked the {\vfour} and found that it indeed matches part of the original text's incorrect spatial information (Fig.~\ref{fig:teaser}(d)). 
Such biased distribution worried the E1 because he did not want such kinds of shortcuts to inflate the model performance, especially when checking the model performance on the dirty set.

\mzhihua{\textbf{Summary.}  Since this shortcut is not an intended solution and inflates model performance, E1 said that he would try to avoid this kind of shortcut in the new dataset construction process, for example, by reducing the number of locality words with \dq{left} (\textbf{R1}, \textbf{R2}, \textbf{R3}). }

\subsubsection{Analyzing Shortcuts without Locality Words}
Then, E1 further explored another shortcut without locality words (Fig.~\ref{Fig.case_study_one_fig_one}(b2)). E1 chose a shortcut that matches sentences where the preposition is \dq{in}, the second word in front of the word is a numeral, and there are no words between these two words (Fig.~\ref{Fig.case_study_one_fig_three}(a), Fig.~\ref{Fig.case_study_one_fig_three}(c)). E1 further checked the What-If Analysis plane and found that the productivity of this shortcut is similar to machine accuracy in the dirty set (Fig.~\ref{Fig.case_study_one_fig_three}(b)). It indicates that the model performance is somehow related to this shortcut. E1 further checked the relationship between shortcuts in the Template View. E1 found that only one child of potential shortcuts matched instances in the test set. This potential shortcut matches sentences where the preposition is \dq{in} and the second word in front of the word is a numeral which is \dq{two} (Fig.~\ref{Fig.case_study_one_fig_three}(d)). 
E1 selected it in the {\vthree} and checked the instances in the {\vfour}. He found that all instances have the same neighborhoods around the highlighted text (Fig.~\ref{Fig.case_study_one_fig_three}(e)). It reminds him that many sentences are modified from the same original sentence. Therefore, when the label distribution is biased for the same original sentence, it allows potential shortcuts to achieve spurious performance on the test set.

\mzhihua{\textbf{Summary.}  Based on the observation that irrelevant text in instances modified from the same original sentence will be matched by shortcuts, it inspired him that when selecting instances in the test set, he should guarantee that the label distribution of the sentences modified from the same original sentence should be more balanced to avoid the potential shortcuts matching irrelevant places in the text (\textbf{R1}, \textbf{R2}, \textbf{R3}).}

\subsection{Case-II: CoLA Dataset}
\begin{figure*}[htb]
\centering 
\includegraphics[width=\linewidth]{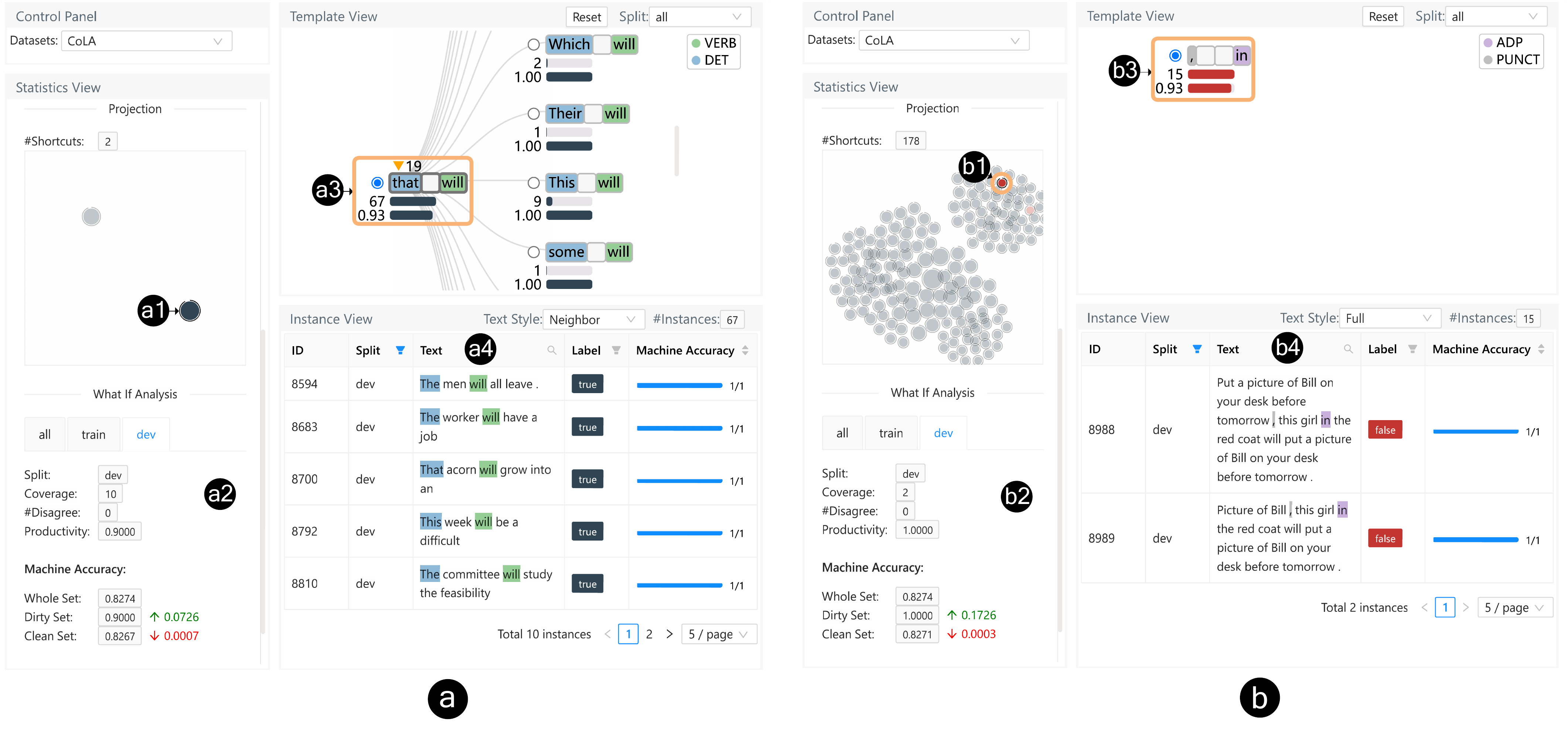}
\caption{E2 selected one of the shortcuts with the prediction label \textit{\dq{true}} in the Projection plane and found that such kinds of irrelevant words can lead to high productivity (a). E2 found one shortcut with the prediction label \textit{\dq{false}} and the instances covered by this shortcut seem to be modified from similar sentences (b). } 
\label{Fig.case_study_two_fig_one}
\end{figure*}

We conducted this case study with a senior NLP researcher (E2). \mzhihua{The background of E2 is depicted in Sec.~\ref{sec:abstractions}.} E2 is working on one project which is related to the grammatical error detection dataset construction. He is very interested in a similar \mzhihua{benchmark dataset} called CoLA~\cite{warstadt2019neural} in the GLUE~\cite{wang2018glue} benchmark. One BERT-based model is finetuned on this benchmark dataset. He is willing to inspect whether there exist some shortcuts in CoLA and gain some insights into avoiding shortcuts in new dataset construction. 

\subsubsection{Analyzing Shortcuts with Prediction Label \textit{\dq{true}}}
E2 selected the CoLA \mzhihua{benchmark dataset} in the Control Panel. 
\mzhihua{He found that the percentage of instances with the Label \textit{\dq{true}} in the development set\footnote[5]{Since the labels of the test set of CoLA are not available, we do not consider the test set in this \mzhihua{benchmark dataset} and assume that the development set serves the role of evaluating the models.} is 0.6913 while the machine accuracy on the development set is 0.8274, which is better than the majority baseline on the development set.}
Since the models achieve high accuracy on the development set, E2 selected the minimum productivity as 0.90 and the minimum coverage as 50. There are only two remaining shortcuts, with the prediction label \textit{\dq{true}} (\textbf{R1}, \textbf{R3}). 

E2 inspected one of them that matches sentences where a verb is \dq{will}, the second word ahead of it is a determiner and similar to \dq{that}, and there is one word between these two words (Fig.~\ref{Fig.case_study_two_fig_one}(a1), Fig.~\ref{Fig.case_study_two_fig_one}(a3)).
\mzhihua{This shortcut covers 10 instances with the same productivity as the machine accuracy (0.9) in the development set (Fig.~\ref{Fig.case_study_two_fig_one}(a2)).} E2 is surprised that such kinds of irrelevant words can lead to high productivity. E2 further inspected the instances in the {\vfour} (Fig.~\ref{Fig.case_study_two_fig_one}(a4)). 
He believed that it is a potential shortcut unintended introduced when collecting data from the biased distribution. Such shortcuts cannot guarantee that the matched sentences are grammatically acceptable. 
For example, the grammatically acceptable sentence \dq{The men will all leave} can be modified as the grammatically unacceptable sentence \dq{The men will all} while both sentences can be matched by this shortcut. 

\mzhihua{\textbf{Summary.} Since the shortcuts are not guaranteed that the sentences can be grammatically correct, E2 thought that negative examples can be constructed from grammatically acceptable sentences by simply removing some critical components of them (\textbf{R2}, \textbf{R3}).}

\subsubsection{Analyzing Shortcuts with Prediction Label \textit{\dq{false}}}
E2 was curious about whether there exist shortcuts with the prediction label \textit{\dq{false}} because instances with the label \textit{\dq{false}} are minor samples compared to instances with the label \textit{\dq{true}}. E2 set the minimum productivity as 0.9 and the minimum coverage as 15. He found that there are only two shortcuts with the same structure and prediction label \textit{\dq{false}} (Fig.~\ref{Fig.case_study_two_fig_one}(b1)). 
This shortcut matches sentences where the first word is punctuation \dq{,}, the second word after the first word is adposition \dq{in}, and there are two words between these two words (Fig.~\ref{Fig.case_study_two_fig_one}(b3)).
E2 further inspected the instances covered by this potential shortcut (Fig.~\ref{Fig.case_study_two_fig_one}(b4)). He found that most sentences before the matched punctuation seem incomplete, then it will potentially lead to grammatically unacceptable sentences as a whole. The punctuation seems a very weak indicator of the incorrect sentence. The two words between the matched text mostly (13/15) are \dq{this girl}. If further considering these two words, all matched sentences are labeled as \textit{\dq{false}}.
In terms of this shortcut, E2 further carefully checked the matched sentences and found that they are mostly modified from similar sentences. But there are only two sentences in the development set (Fig.~\ref{Fig.case_study_two_fig_one}(b2)). 

\mzhihua{\textbf{Summary.} Based on the observation that there are a small number of instances matched by this shortcut.
E2 thought that ignoring this shortcut seems not too problematic. E2 said that to solve this shortcut, he can take a simple fix like adding one more positive sentence which can be matched by this shortcut in the development set (\textbf{R2}). }

\subsection{Expert Interviews}
\mzhihua{We recruited eight NLU experts (E3-E10, age: 23-50, 5 males, 3 females) by email and then conducted interviews with them to collect their feedback on the usefulness and usability of the system.} They all have more than two years of experience in NLU dataset construction.
E3 has participated in constructing various kinds of NLU datasets such as treebank, semantic roles labeling, spatial reasoning, and so on. E4 has proposed two NLP-related datasets. E6 mainly focuses on dataset quality evaluation during the NLU dataset construction process. E8 focuses on premise-based multimodal reasoning dataset construction. E3, E5, and E7 have worked on constructing the SpaCE2021 \mzhihua{benchmark dataset} and are going to build a better dataset to evaluate the spatial reasoning capability of NLU models. \mzhihua{E9 has worked on dataset construction for error detection for the generated text from pre-trained language models. E10 has worked on constructing an adversarial dataset for the GLUE benchmark.}
\mzhihua{None of them (E3-E10) are our co-authors nor tried the system (\name) before the interviews. }

For the procedures of the expert interviews, we first introduced {\name} to them, including the shortcut mining algorithm and interface, and showed the experts the system usage using case one (in Sec.~\ref{sec:case_one}). Then we asked them to follow the workflow in the demonstrated case study to explore the system using the SpaCE2021 dataset.  
When exploring the system, they were asked to identify potential shortcuts in the \mzhihua{benchmark dataset} and figure out actionable insights to fix the shortcuts. They can freely ask questions and provide comments and suggestions on the system. We summarize the results of expert interviews in the following paragraphs.

\textbf{Shortcut exploration.}
All experts confirmed that {\name} could facilitate systematic and comprehensive shortcut exploration and inspire them to take action to fix shortcuts. 
Compared to single gram analysis in their previously-used workflows, experts found that our system helps generalize their findings of shortcuts by considering different relationships (e.g., structural and semantic similarities) between data instances.
\mzhihua{For example, E3 said that previously he found that sentences with the words \dq{palm left} are easy for models to solve. By checking the relationship between the shortcuts, it is interesting to find that the sentences where a locality word “left" is following a noun word are mostly labeled as \textit{\dq{false}}.}
\mzhihua{Moreover, E5 and E9 mentioned that associating the shortcuts with data instances assists them in identifying undesirable shortcuts.
For example, E9 observed that there is one shortcut which matches sentences with the word \dq{top}. E9 checked the matched text for this shortcut. E9 found that most of the text is highly related to grammatical errors and should be fixed.}
In addition, what if analysis supported by the system were considered useful to estimate the severity of shortcuts and identify those that need immediate actions for fixing.
\mzhihua{For example, E4 noticed that there is one shortcut which matches sentences where the word \dq{side}. The machine performance of the dirty set of the shortcut is even higher than the productivity of the shortcut. It means that this shortcut is exploited by the models. 
E4 thought that it is urgent to fix this shortcut.}

\textbf{Visualization and interaction designs.}
Overall, experts agreed that visualizations and system interactions are useful and easy to learn and use. 
Particularly, experts favored the {\vthree} for summarizing a group of potential shortcuts with relatively simpler representations.
\mzhihua{E10 said that \textit{\dq{the potential shortcuts are somehow hard to understand in the beginning, but the glyph of them seems simple and intuitive to understand.}}
However, E3 noticed that he sometimes needs to scroll long lists of templates in the {\vthree}.}
Experts also appreciated the {\vfour}, which provides detailed context information about the data instances covered by potential shortcuts. The highlighting of components of potential shortcuts helps them quickly reason shortcut patterns with concrete example sentences.
\mzhihua{For the {\vtwo}, E7 and E10 mentioned sometimes he needs to adjust filtering to reduce visual clutter. Besides the general statistical results. E4 desired more descriptions of the glyphs in the system.}

\textbf{Suggestions for improvement.}
Experts were generally eager to use the system in their future work to inspect the newly constructed dataset and find and mitigate potential shortcuts in them.
Nevertheless, they identified some system limitations and provided several valuable suggestions. 
\mzhihua{E5, E8, E9, and E10 requested system support for editing and improving datasets (besides what-if analysis) after discovering undesirable shortcuts. For example, E9 commented that \textit{\dq{for those texts with shortcuts including spatial words, we can use pre-trained language models to fix it, like picking a spatial word generated by pre-trained language models.}}}
E6 stated that it would be better to incorporate a higher-level shortcuts mining algorithm in the system. \mzhihua{For example, incorporating more types of components, like named entities, dependency, and logical structure, in shortcuts will be interesting.}
In addition, other system functions were mentioned. For example, E7 suggested that the system should provide a function to switch between different languages to accommodate users from different countries. E3 and E4 asked us to provide more documentation and tutorials on the system to help them set up the system for their customized data. 
\section{Discussion and Future Work}


In this section, we will summarize the lessons learned during the design and evaluation of {\name}. We will also discuss the limitations and future work.

\subsection{Lessons Learned}

\mzhihua{\textbf{Contextualizing shortcuts with concrete instances.}
To present an overview of data instances covered by potential shortcuts, we have proposed glyphs to provide the abstraction of shortcuts using templates based on POS, relative positions between words, and semantic meaning. 
During expert interviews, the experts appreciated the compactness and intuitiveness of the visual designs, especially templates for summarizing shortcuts. Then, to decide whether the potential shortcuts are undesirable or not, they would like to see the context of those shortcuts in original sentences in the Instance View.
Such information can help them understand the reason and logic behind the emergence of specific shortcuts. 
Therefore, we expect future visual analytics systems should tightly associate the abstract summary with detailed context to facilitate a deep and comprehensive understanding of complex concepts.}

\mzhihua{\textbf{Focusing on data issues besides model issues.} 
Recent machine learning models, especially deep learning models for NLP, achieve compelling performance on different benchmark datasets. However, they are far from being perfect since their performance could be inflated by dataset biases (e.g., shortcuts). 
In this paper, we design and build {\name} to help dataset creators systematically explore shortcuts in NLU datasets and conduct what-if analyses. The case studies and expert interviews confirm that our system can help users discover shortcuts and gain valuable insights into mitigating them in the dataset construction. 
We hope that our work can draw attention to data issues in the AI community.}

\subsection{Limitations and Future Work}
\mzhihua{\textbf{Investigate more complex shortcuts in NLU benchmark datasets.}
In this paper, we extract and visualize shortcuts considering POS, relative positions between words, and semantic meaning.
In the future, we can further consider other linguistic properties of shortcuts, such as the named entities, dependency, and logical structure. 
Moreover, we adopt matching-based and similarity-based methods to decide the coverage of shortcuts. That is, if a data instance is covered by a shortcut, it must satisfy or resemble the condition of the shortcut.
However, other types of decision conditions can be studied. For example, exclusion-based shortcuts cover instances that do not contain the components of the specific shortcuts.
And count-based shortcuts cover the instances that contain specific components multiple times.
}

\mzhihua{\textbf{Support shortcut exploration for more diverse NLU tasks and benchmark datasets.} 
In this paper, we demonstrate our system for shortcut exploration through two single sentence classification tasks, including grammatical acceptability judgment and spatial reasoning.
We can further extend the scope of our system to support multiple sentence classification tasks (e.g., natural language inference). For example, we can investigate the co-occurrence of our defined shortcuts in multiple sentences.
Moreover, it will be beneficial to improve the system to investigate shortcuts in text generation tasks, \mzhihua{such as question answering~\cite{rogers2021qa} and story generation~\cite{guan2020knowledge}.}}
Last but not least, the current system supports analyzing \mzhihua{benchmark datasets} in Chinese (i.e., SpaCE2021) and English (i.e., CoLA). It is valuable for our system to support \mzhihua{benchmark datasets} in different languages.

\mzhihua{\textbf{Enable interactive shortcuts fixing.} 
Currently, {\name} facilitates multi-level exploration of shortcuts regarding productivity, coverage, and semantic and structural information. 
It helps experts discover shortcuts to be further fixed. Also, the system provides the what-if analysis to estimate the influences of removing potential shortcuts in the datasets. 
As suggested by the experts during the interview in the evaluation, we can further enhance the system usability with more choices to fix shortcuts. 
For example, we can enable users to add or modify instances to adjust the distribution of labels using adversarial text generation~\cite{zang2019word, zeng2020openattack, jin2020bert} and model-based text generation methods~\cite{schick2021generating,zhao2021lmturk, liu2022wanli }. 
Besides, it is possible that when we solve one shortcut, another shortcut will appear. 
One possible solution is that we can visualize the label distribution shift of other related potential shortcuts in the system so that users understand the possible consequences and trade-offs of fixing shortcuts of interest. \mzhihua{Furthermore, we can use statistics of shortcuts and machine performance to measure the quality of the dataset. Users can observe the benefits gained from their actions for improving the dataset.} }

\textbf{Improve scalability of visual designs to handle a larger number of instances and classes.} 
The current visual designs cannot scale well when the number of data instances and classes increases. For example, the Projection plane in the {\vtwo} only supports up to 300 shortcuts. To enable the functionality (e.g., data selection) of the Projection plane, users need to set the minimum coverage and productivity to filter the potential shortcuts. Moreover, in the {\vthree}, if the number of selected shortcuts is large, users may need to scroll down for a while to find shortcuts of their interest. It would be useful to have a search function for users to look for shortcuts containing words and structures of their interest. \mzhihua{Last but not least, since we use color to encode labels, if the number of categories is large (i.e., 100), different colors for different classes may be indistinguishable. The system should also support displaying the labels near the glyph of shortcuts if this situation occurs.}

\section{Conclusion}
NLU benchmarks are found to suffer the spurious bias and inflate the model performance in recent years. To further help NLU experts to build challenging and pertinent \mzhihua{benchmark datasets,} in this paper, we develop a system called {\name} to support dataset creators in conducting multi-level exploration of shortcuts in the \mzhihua{NLU benchmark dataset.} Specifically, the {\vtwo} helps users grasp the overview of shortcuts in the \mzhihua{benchmark dataset} and conduct a what-if analysis of shortcuts of interest. The {\vthree} employs hierarchical and interpretable templates to help summarize the \mzhihua{benchmark dataset} issues through shortcuts. The {\vfour} allows users to further explore the instances covered by shortcuts. We have conducted two case studies with the experts on understanding and mitigating shortcuts in \mzhihua{NLU benchmark datasets,} as well as expert interviews with eight experts to demonstrate the usefulness and usability of the {\name}.
\bibliographystyle{IEEEtran}
\bibliography{main}

%

\begin{IEEEbiography}[{\includegraphics[width=1in,height=1.25in,clip,keepaspectratio]{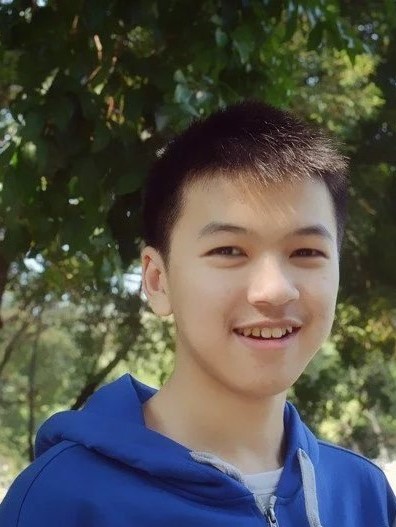}}]{Zhihua Jin} is currently a Ph.D. student at the Hong Kong University of Science and Technology (HKUST). He received his BEng degree in Computer Science and Technology from Zhejiang University in 2019. His research interests lie in the intersection of visualization and machine learning, especially explainable artificial intelligence (XAI). 
\end{IEEEbiography}

\begin{IEEEbiography}[{\includegraphics[width=1in,height=1.25in,clip,keepaspectratio]{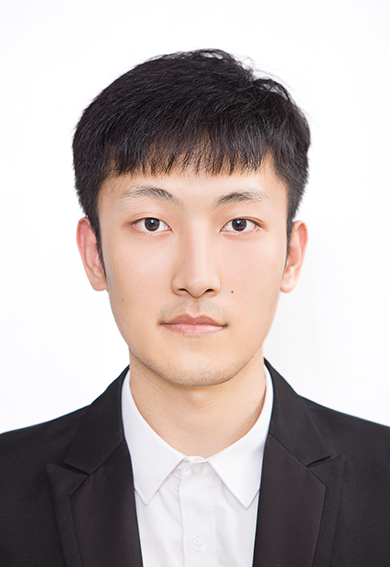}}]{Xingbo Wang} is a Ph.D. candidate in the Department of Computer Science and Engineering at the Hong Kong University of Science and Technology (HKUST). He obtained a B.E. degree from Wuhan University, China in 2018. His research interests include human-computer interaction (HCI), data visualization, natural language processing (NLP), and multimodal analysis. For more details, please refer to \url{https://andy-xingbowang.com/}.
\end{IEEEbiography}


\begin{IEEEbiography}[{\includegraphics[width=1in,height=1.25in,clip,keepaspectratio]{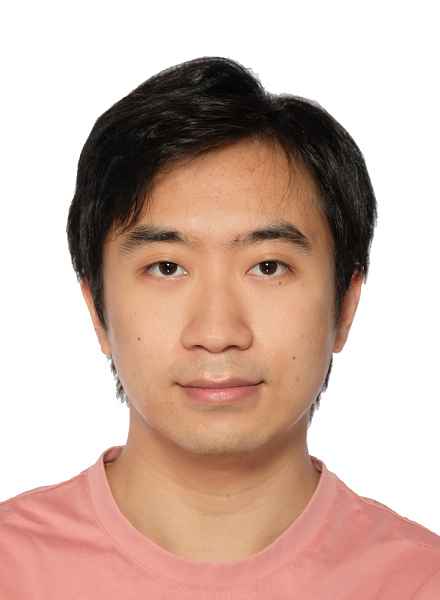}}]{Furui Cheng} is a Ph.D. candidate in the Department of Computer Science and Engineering at the Hong Kong University of Science and Technology (HKUST). He obtained a B.E. degree from Beihang University, China, in 2018. His research interests include visual analytics, eXplainable AI (XAI), and biomedical AI. 
\end{IEEEbiography}

\begin{IEEEbiography}[{\includegraphics[width=1in,height=1.25in,clip,keepaspectratio]{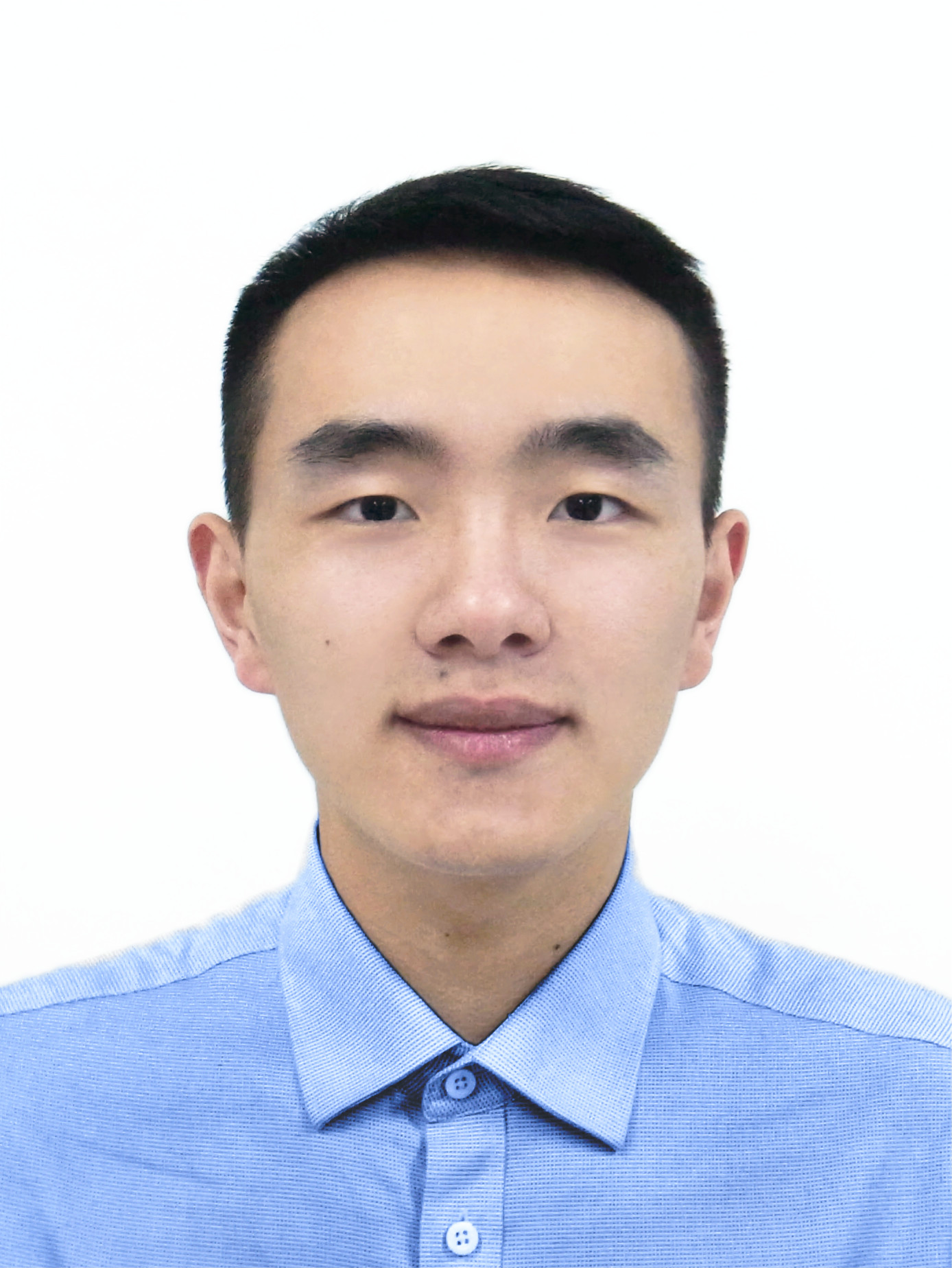}}]{Chunhui Sun} is a Ph.D. candidate in the Department of Chinese Language and Literature at Peking University (PKU). He obtained a M.Ed. degree from Soochow University, China, in 2018. His research interests lie in the areas of Computational Linguistics, Cognitive Linguistics, Natural Language Understanding Benchmarks, and eXplainable AI (XAI), etc.
\end{IEEEbiography}

\begin{IEEEbiography}[{\includegraphics[width=1in,height=1.25in,clip,keepaspectratio]{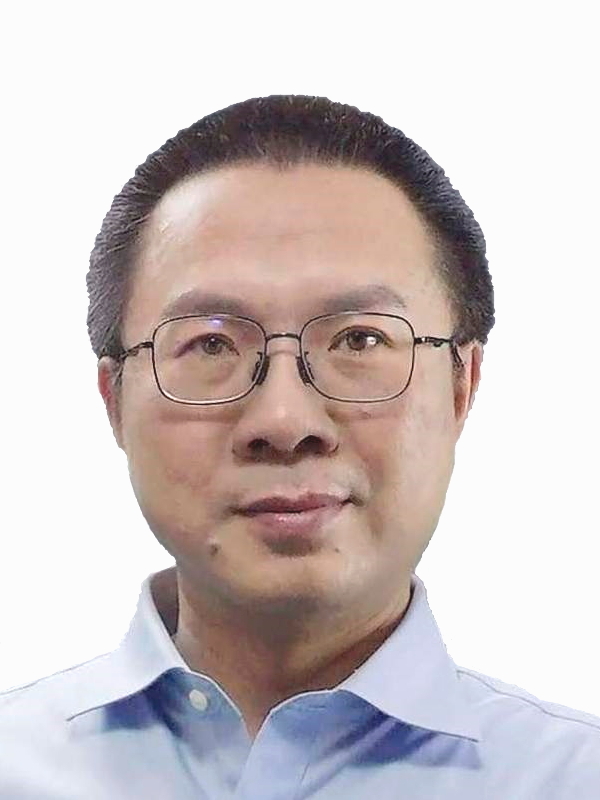}}]{Qun Liu} is the Chief Scientist of Speech and Natural Language Processing of Huawei Noah's Ark Lab. He was a Full Professor in Dublin City University and the Theme Leader of the ADAPT Centre, Ireland during July 2012 and June 2018. Before that, he was as a Professor in the Institute of Computing Technology (ICT), Chinese Academy of Sciences for 20 years, where he founded and led the ICT NLP Research Group. He obtained a B.Sc. in computer science in the University of Science and Technology of China, a M.Sc. in Chinese Academy of Sciences, and a Ph.D. in Peking University. His research interests lie in the areas of Natural Language Processing, Machine Translation, Pre-trained Language Models, etc.
\end{IEEEbiography}

\begin{IEEEbiography}[{\includegraphics[width=1in,height=1.25in,clip,keepaspectratio]{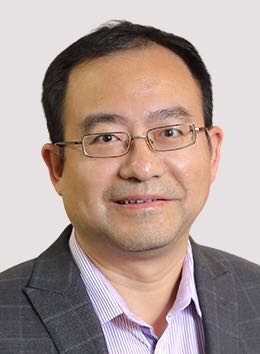}}]{Huamin Qu} is a professor in the Department of Computer Science and Engineering (CSE) at the Hong Kong University of Science and Technology (HKUST) and also the director of the interdisciplinary program office (IPO) of HKUST. He obtained a BS in Mathematics from Xi'an Jiaotong University, China, an MS and a PhD in Computer Science from the Stony Brook University. His main research interests are in visualization and human-computer interaction, with focuses on urban informatics, social network analysis, E-learning, text visualization, and explainable artificial intelligence (XAI).
\end{IEEEbiography}



\appendices

\section{Actionable Insights Verification}
To verify actionable insights in case studies, we can use two ways to evaluate whether the dataset's quality has improved after taking action to fix shortcuts. 
The first way is to inspect the model performance before and after fixing shortcuts in the test dataset.
If the model performance on the new test set is less than the model performance on the original test set, it indicates that the dataset has become challenging. 
The second way is to inspect the number of shortcuts before and after fixing shortcuts under certain filtering conditions. 
The shortcut mining algorithm will be run on the modified dataset to ensure that all potential shortcuts defined in the paper are taken into account. 
Suppose the number of shortcuts decreases and the target shortcut disappears under the original filtering conditions. 
In that case, it indicates that the shortcut has been fixed, and no more shortcuts are added to the dataset. 
That is, after taking action to fix the shortcut, many other shortcuts will also disappear. Another situation is that some shortcuts disappear, and new shortcuts appear. 
Since the number of shortcuts decreases, it indicates that the number of shortcuts disappearing is larger than the number of  shortcuts appearing. 
When experts determine that the shortcuts should be fixed, it indicates that they are the source of task-irrelevant cues in the dataset. 
In other words, removing the non-pertinent instances or decreasing the number of shortcuts can bring a more pertinent dataset, improving the dataset quality. 

We acknowledge the following limitations:
1) We consider only a limited number of models in the system. For models outside our scope, even if the model performance in the system decreases for a clean set, it is not guaranteed that the performance of the models outside our scope will decrease.
2) When adding or modifying instances in the dataset, it is necessary to know the model prediction results for the new instances. If the model inference is not available, the performance of models on the new dataset cannot be accurately calculated.
3) In terms of inspecting the number of shortcuts in the dataset, we only consider the potential shortcuts defined in the paper, which can be extracted using the shortcut mining algorithm. There are also other kinds of shortcuts that are outside of our scope, which also cannot be revealed by our shortcut mining algorithm. 
For those shortcuts, we cannot detect whether those shortcuts exist in the dataset via the current shortcut mining algorithm. 
Consequently, we cannot detect whether new shortcuts outside of our scope have been introduced into the dataset after we take action to fix one shortcut in the dataset.

In the following paragraphs, we report the verification results of actionable insights in case studies.

\subsection{Case-I: SpaCE2021}
When we set the minimum productivity as 0.75 and minimum coverage as 10, the number of shortcuts is 232. The coverage of those shortcuts is 308. If removing those instances and consequently removing the shortcuts, the machine performance on the clean set is 0.6417 (0.0507 less than the machine performance on the whole test set, which is 0.6924). It implies that if the experts investigate all those shortcuts and fix them, the machine performance can drop a lot and will make the dataset more challenging.

\subsubsection{Fixing Shortcuts with Locality Words}

In terms of improving this dataset and fixing shortcuts with locality words, E1 suggested that \textit{we can remove those instances which are covered by the target shortcut with the locality word \dq{left} and noun word ahead in this dataset}. Following E1's suggestions, we removed 15 instances which are covered by this target shortcut in the test set of SpaCE2021. After rerunning the shortcut mining algorithm in the new dataset and reloading the data into the system, we set the minimum productivity as 0.75 and the minimum coverage as 100. Compared to the original dataset, we found that two shortcuts disappear. One shortcut is the target shortcut. Another shortcut is matching the sentences with the locality word \dq{right} and a verb which is two words behind the locality word \dq{right}.
We further set the minimum productivity as 0.75 and minimum coverage as 10. Compared to the number of shortcuts in the original dataset, which is 232, the number of shortcuts in the new dataset is 211 (21 less than the number of shortcuts in the original dataset). The machine performance in the new dataset is 0.6872 (0.0052 less than the machine performance in the original dataset). Those results imply that this action can make the dataset more challenging and pertinent.

\subsubsection{Fixing Shortcuts without Locality Words}
In terms of improving this dataset and fixing shortcuts without locality words, E1 suggested that \textit{we can remove some instances where the preposition is \dq{in}, the second word in front of the word is a numeral which is \dq{two}, and the label is \textit{\dq{false}} to balance the label distribution of the sentences
modified from the same original sentence.}
Following E1's suggestions, we removed 15 instances with the label \textit{\dq{false}} to balance the label distribution of the sentences covered by the target shortcut. After rerunning the shortcut mining algorithm and reloading the data into the system, we set minimum productivity as 0.75 and minimum coverage as 100. The target shortcut has been removed. When we set the minimum productivity as 0.75 and the minimum coverage as 10, the number of shortcuts in the original dataset is 232 while the number of shortcuts in the new dataset is 224 (8 less than the number of shortcuts in the original datasets). The machine performance in the new dataset is 0.6882 (0.0042 less than the machine performance in the original dataset). Those results imply that this action can improve dataset quality.

\subsection{Case-II: CoLA}
When we set the minimum productivity as 0.75 and minimum coverage as 10, the number of shortcuts is 754. The coverage of those shortcuts is 382. If removing those instances and consequently removing the shortcuts, the machine performance on the clean set is 0.7595 (0.0679 less than the machine performance on the whole test set, which is 0.8274). It implies that if the experts investigate all those shortcuts and fix them, the machine performance can drop a lot and will make the dataset more challenging.

\subsubsection{Fixing Shortcuts with Prediction Label \dq{true}}
To solve this target shortcut which matches sentences where a verb is \dq{will}, the second word ahead of it is a determiner and similar to \dq{that}, and there is one word between these two words, E2 suggested that \textit{we can add negative sentences which can be matched by this shortcut in the
development set via removing some critical components of grammatically acceptable sentences.}
Following E2's suggestions, we add two new sentences with the label as \dq{false}. One sentence is \dq{The men will all.} Another sentence is \dq{This week will a difficult one for us.} They can be matched by the target shortcut. After rerunning the shortcut mining algorithm and reloading the data into the system, we set the minimum productivity as 0.9 and the minimum coverage as 50. We found that the target shortcut has been removed. We further set the minimum productivity as 0.75 and the minimum coverage as 10. The number of shortcuts in the original dataset is 754, and the number of shortcuts in the new dataset is 750 (4 less than the number of shortcuts in the original dataset). The machine performance in the new dataset is 0.8258 (0.0016 less than the machine performance in the original dataset). The results demonstrate that adding negative sentences can improve dataset quality.

\subsubsection{Fixing Shortcuts with Prediction Label \dq{false}}
To solve this target shortcut which matches sentences where the first word is punctuation
\dq{,}, the second word after the first word is adposition \dq{in}, and
there are two words between these two words, E2 suggested that \textit{we can add one grammatically acceptable sentence which can be matched by this shortcut in the
development set.}
Following E2's suggestions, we add one sentence which is \dq{When we put a picture of Bill on your desk before tomorrow, this girl in the red coat will put a picture of Bill on your desk before tomorrow.} The label of this sentence is \textit{\dq{true}}. The target shortcut can match this sentence.
After rerunning the shortcut mining algorithm and reloading the data into the system, we set the minimum productivity as 0.9 and the minimum coverage as 15. We found that two target shortcuts are removed. We further set the minimum productivity as 0.75 and the minimum coverage as 10. The number of shortcuts in the original dataset is 754, and the number of shortcuts in the new dataset is 749 (5 less than the number of shortcuts in the original dataset). The machine performance in the new dataset is 0.8266 (0.0008 less than the machine performance in the original dataset). The results demonstrate that constructing a positive sentence can improve dataset quality.

\end{document}